\documentclass[sigconf]{acmart}
\AtBeginDocument{%
  }

\usepackage{soul}
\usepackage{xcolor}
\sethlcolor{yellow}
\usepackage{relsize}

\usepackage{tabularx} 
\usepackage{booktabs} 
\usepackage{array}    

\usepackage{soulutf8}
\sethlcolor{yellow!20}
\soulregister{\cite}{1}
\soulregister{\citep}{1}
\soulregister{\citet}{1}

\let\oldtexttt\texttt
\renewcommand{\texttt}[1]{{\small\oldtexttt{#1}}}

\copyrightyear{2025}
\acmYear{2025}
\setcopyright{cc}
\setcctype{by}
\acmConference[ASSETS '25]{The 27th International ACM SIGACCESS Conference on Computers and Accessibility}{October 26--29, 2025}{Denver, CO, USA}
\acmBooktitle{The 27th International ACM SIGACCESS Conference on Computers and Accessibility (ASSETS '25), October 26--29, 2025, Denver, CO, USA}
\acmDOI{10.1145/3663547.3746346}
\acmISBN{979-8-4007-0676-9/2025/10}
\begin{document}

\title[Adapting Non-Speech Captions with Anchored Generative Models]{CapTune: Adapting Non-Speech Captions With Anchored Generative Models}

\author{Jeremy Zhengqi Huang}
\affiliation{%
    \institution{University of Michigan}
    \city{Ann Arbor, MI}
    \country{USA}}
\email{zjhuang@umich.edu}

\author{Caluā de Lacerda Pataca}
\affiliation{%
    \institution{Rochester Institute of Technology}
    \city{Rochester, NY}
    \country{USA}}
\email{cd4610@rit.edu}

\author{Liang-Yuan Wu}
\affiliation{%
    \institution{University of Michigan}
    \city{Ann Arbor, MI}
    \country{USA}}
\email{lyuanwu@umich.edu}

\author{Dhruv Jain}
\affiliation{%
    \institution{University of Michigan}
    \city{Ann Arbor, MI}
    \country{USA}}
\email{profdj@umich.edu}

\renewcommand{\shortauthors}{Huang et al.}

\begin{abstract}    
    Non-speech captions are essential to the video experience of deaf and hard of hearing (DHH) viewers, yet conventional approaches often overlook the diversity of their preferences. We present CapTune, a system that enables customization of non-speech captions based on DHH viewers’ needs while preserving creator intent. CapTune allows caption authors to define safe transformation spaces using concrete examples and empowers viewers to personalize captions across four dimensions: level of detail, expressiveness, sound representation method, and genre alignment. Evaluations with seven caption creators and twelve DHH participants showed that CapTune supported creators’ creative control while enhancing viewers’ emotional engagement with content. Our findings also reveal trade-offs between information richness and cognitive load, tensions between interpretive and descriptive representations of sound, and the context-dependent nature of caption preferences.
\end{abstract}



\begin{teaserfigure}
  \centering
  \includegraphics[width=0.98\textwidth]{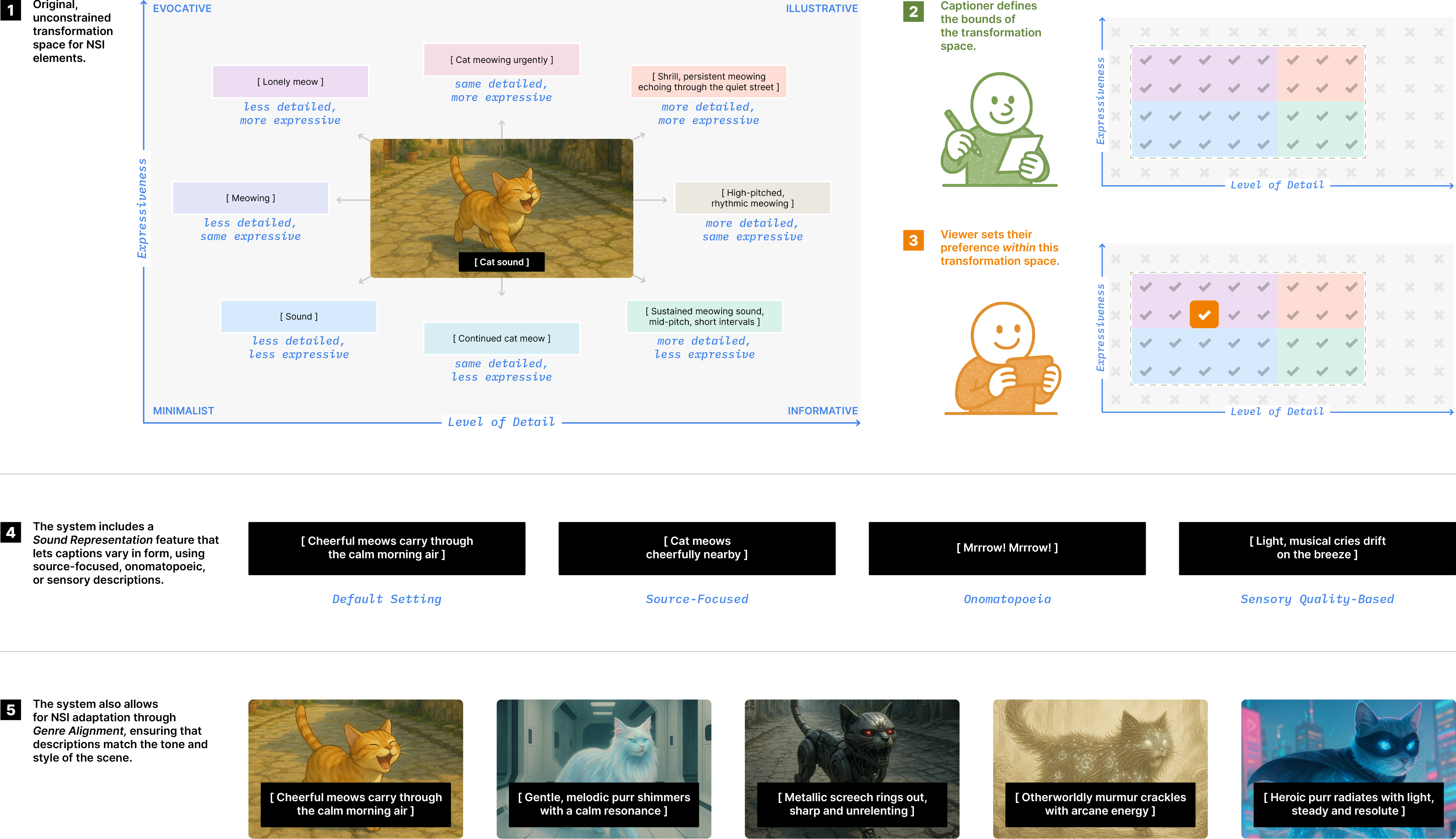}
  \caption{Overview of the CapTune transformation system. (1) Non-speech caption transformations are mapped within a two-dimensional space defined by \textsc{Level of Detail} and \textsc{Expressiveness}. (2) Captioners define the bounds of this space for a given caption file. (3) Viewers select their preferences within the defined range. (4) Captions can be transformed using one of three \emph{sound representation} methods: source-focused, onomatopoeia, or sensory quality-based. (5) Captions can also be \emph{aligned with genre} tone and style for narrative~coherence.}
  \Description{A visual diagram illustrating the CapTune system for customizing non-speech captions. The left section shows a 2D grid with axes labeled "Expressiveness" (vertical) and "Level of Detail" (horizontal). Various caption styles for a cat meowing are plotted on this grid, ranging from minimal and vague ([Sound]) to rich and evocative ([Shrill, persistent meowing echoing through the quiet street]). At the center of the grid is an image of an orange cartoon cat. The right section includes two smaller diagrams: one showing a captioner defining the editable region within the grid (green figure), and another showing a viewer selecting their preferred caption style within that space (orange figure). Below, two rows illustrate (1) different captioning styles using the same cat sound — default, source-focused, onomatopoeic, and sensory-based — and (2) examples of genre-aligned caption variations for different visual scenes, including a cheerful animation, a calm sci-fi scene, a dark fantasy scene, and a heroic sci-fi moment.}
  \label{fig:teaser}
\end{teaserfigure}

\maketitle

\section{Introduction}
Non-speech information (NSI) in captions, including sound events and effects (e.g., \texttt{[Soft Piano music]}, \texttt{[Thunder Rumbling]}) and extra-speech information (e.g., tones of speech), is a crucial component of media accessibility, enabling deaf and hard of hearing\footnote{We use the term ``deaf and hard of hearing'' to represent diverse perspectives, including deaf, Deaf, hoh, black-deaf, or geographic deaf communities.} (DHH) viewers to access auditory cues in video content. However, conventional NSI captioning systems face a fundamental limitation: they adopt a ``one-size-fits-all'' approach, where creators provide a single, fixed caption track that may fail to account for the diversity in viewers' backgrounds and preferences. Some DHH viewers prefer richly descriptive, creative captions that convey subtle acoustic nuance, while others favor more concise, minimal captions focused on core information \cite{samwellbarley_deaf_2021, great-typhoon_what_2023}. For example, in an atmospheric thriller, a generic caption like \texttt{[Ominous Music]} may undersell the dramatic tension, while in a nature documentary, an overly figurative caption such as \texttt{[Furious Gusts Rip Through the Forest]} may feel distracting and inappropriate. Cultural and regional differences within DHH communities also shape how sound is interpreted and valued \cite{ladd_understanding_2008, friedner_sound_2012}.

These nuances underscore a pressing need for preference-driven captioning approaches that allow viewers to adapt captions to their personal goals and preferences---whether clarity, immersion, or efficiency---and align them with the tone and style of the media. While recent work has explored visual augmentations of NSI, such as font styles \cite{de_lacerda_pataca_caption_2024}, color schemes \cite{lee_emotive_2007}, and visual overlays (e.g., emojis and visualizations) \cite{may_enhancing_2023, wang_visualizing_2017, alonzo_beyond_2022}, these primarily focus on how captions look, rather than the semantic content. Text remains the most common and accessible medium for NSI in formal media, such as movies. Moreover, stylized NSI captions in popular media (e.g., Stranger Things \cite{salazar_wet_2022, alepa_stranger_2022, bitran_meet_2022}) suggest growing recognition for their expressive potential; however, these are still static and non-customizable.


To explore alternatives, we conducted a qualitative analysis of DHH viewers' first-person accounts from online communities to understand their current experiences and needs regarding NSI captions. This analysis identified four key design opportunities for customizable caption experiences: (1) varying levels of detail for narrative clarity, (2) expressiveness to support emotional and narrative resonance, (3) personalizable representation methods for sound, and (4) genre and style alignment for thematic coherence.

We implemented these design opportunities through CapTune, a system that leverages generative models (e.g., large language models) to support viewer-driven customization of non-speech captions while preserving creator intent. The system comprises two components: \textit{Creator Tool}, which allows creators to define transformation boundaries using two anchor points along key parameters---\textsc{Level of Detail} and \textsc{Expressiveness}---providing guardrails for generative transformations; and \textit{Viewer Client}, which adapts captions during playback based on user-specified preferences. By combining creator-defined constraints with viewer-controlled customization, CapTune introduces a workflow for co-authored, adaptive caption experiences.

We evaluated CapTune through user studies with seven creators and twelve DHH participants, assessing usability, support for narrative engagement, and perceived quality of the adapted captions. Results showed that CapTune enhanced DHH viewers' emotional and narrative engagement with content while supporting creators' editorial control. Participants also emphasized the importance of adapting caption style based on genre, scene context, and viewing intent, highlighting the need for flexible, context-aware customization in accessible media.

In summary, our contributions are threefold:
\begin{enumerate}
    \item We offer insights into nuanced captioning preferences of DHH viewers, drawn from the analysis of online DHH community discussions.
    \item We introduce CapTune, a system for personalized transformation of NSI captions using creator-defined constraints and viewer-controlled parameters.
    \item We present empirical insights from evaluations of 7 creators and 12 DHH participants, and distill our findings into concrete design recommendations for future captioning systems that respect creator intent while enabling viewers to adapt non-speech information to their own viewing goals, media contexts, and access needs.
\end{enumerate}



\section{Related Work}
We provide background on DHH culture and sound perception, situating our work within the captioning of non-speech/paralinguistic cues, as well as AI-driven systems for DHH users.

\subsection{DHH Culture and DHH People's Perception of Sound}
The DHH community encompasses distinct social norms, beliefs, traditions, and values. Deafness is traditionally understood through several models that shape perceptions and practices around DHH individuals: medical, social, and cultural-linguistic. The medical model frames hearing loss as a physical condition requiring intervention \cite{lane_deaf_2002, noauthor_medical_2013}. The social model emphasizes societal barriers that restrict full participation by DHH individuals \cite{national_research_council_us_committee_on_disability_determination_for_individuals_with_hearing_impairments_hearing_2004, noauthor_medical_2013}. In contrast, the cultural-linguistic model offers a more nuanced understanding by highlighting the shared experiences, languages, and values of DHH communities \cite{ladd_understanding_2003, holcomb_introduction_2013}. Central to this model is the use of sign languages, which serve not only as a primary communication modality but also as a cornerstone of cultural identity \cite{noauthor_what_2021}. Over centuries, sign languages such as American Sign Language (ASL) have evolved rich phonological, morphological, and syntactic structures capable of expressing complex objects, emotions, and narratives \cite{sandler_sign_2006, liddell_grammar_2003, valli_linguistics_2011}.

Creating cinematic captions that resonate with DHH audiences requires a deep understanding of how DHH people perceive and interpret sound. A review of American Deaf literature revealed that Deaf writers often represent sound through alternate modalities (e.g., vision, tactile) and emotional and cognitive associations (e.g., sense of rhythm, feelings) \cite{rosen_representations_2007}. Recently, Deaf artist Christine Sun Kim illustrated how she would ``rewrite the closed captions,'' offering a first-person account of sound perception that closely parallels how ASL perceives sounds while also incorporating personal and affective nuance \cite{pop-up_magazine_artist_2020}. These perspectives underscore the importance of \emph{cultural adaptation} as a core dimension in transforming cinematic captions.

\subsection{Incorporating Non-Speech Information and Paralinguistic Cues in Video Captions}
Closed captioning is a core component of accessible video experiences for DHH individuals. Yet, until recently, the captioning industry has primarily focused on speech, often overlooking non-speech information (NSI; e.g., sound effects, music) and paralinguistic cues (e.g., tone, volume, rate) \cite{downey_closed_2008}. While current guidelines do include NSI specifications---such as formatting and timing \cite{noauthor_captioning_2025, noauthor_english_2025, initiative_wai_captionssubtitles_2025}---these elements are often treated as technical or legal requirements rather than opportunities for enhancing narrative experience \cite{zdenek_reading_2015}. Beyond functional accuracy, how captions convey NSI and paralinguistic cues in ways that align with creative intent and narrative remains an open challenge. Prior formative research has shown that DHH viewers’ preferences for such cues vary widely, shaped by factors such as genre, plot relevance, and cultural background \cite{may_enhancing_2023, alonzo_beyond_2022}. For example, many prefer textual over graphical captions for longer or more serious productions (e.g., movies) \cite{alonzo_beyond_2022}.

To date, most caption enhancements for NSI and paralinguistic cues rely on visual augmentation---either typographic (e.g., font color, weight, or size) or graphic overlays (e.g., waveforms, emojis, or animations). For instance, \citet{de_lacerda_pataca_caption_2024} explored how affective properties of speech (e.g., valence and arousal) can be conveyed through typography (e.g., font color). Meanwhile, Android recently deployed “Expressive Captions” that emphasize emotional salience through capitalization \cite{noauthor_expressive_2024}. Visual overlays such as animated text \cite{wang_visualizing_2017}, emoticons \cite{lee_emotive_2007}, and waveform animations \cite{may_enhancing_2023} also aim to enrich caption presentation. May et al. \cite{may_towards_2024} proposed embedding metadata, such as speaker position or mood, to enable user-driven caption customization.

Building on these efforts, our work extends prior findings through an online probe of DHH viewers' preferences for cinematic captioning, with a focus on production-scale, story-driven content such as movies and documentaries. Importantly, we also consider caption writers as key stakeholders---whose creative decisions and workflow constraints shape how NSI is represented---motivated by growing tensions between accessibility and artistic expression \cite{salazar_wet_2022}. Moreover, we address a core limitation of current captioning systems: their “one-size-fits-all” approach to NSI and paralinguistic cues. Rather than relying solely on visual styling, we propose a caption customization pipeline that transforms the textual content itself to adapt to the diverse, personalized needs of DHH viewers.

\subsection{AI-Driven Systems for DHH Users}
While the DHH community has long benefited from advances in machine learning and natural language processing, adapting these systems to their diverse needs remains an ongoing challenge. Recent HCI research has explored AI-powered sound awareness technologies, leveraging deep learning-based sound classification to build home, mobile, and wearable systems tailored to user preferences \cite{bragg_personalizable_2016, jain_homesound_2020, jain_soundwatch_2020}. More recent systems have emphasized a symbiotic relationship between user and system, supporting personalization through user-supplied training data \cite{jain_protosound_2022} or reinforced feedback \cite{do_adaptivesound_2023}. SoundWeaver \cite{huang_weaving_2025}, for example, introduced an \textit{intent-driven} framework for aligning AI behavior with users’ situational goals.

These efforts reflect a broader trend toward aligning AI outputs with user intent. However, few have addressed narrative media, where sound is interpretive, layered, and emotionally charged. CapTune expands this space by focusing on personalized captions for cinematic experiences. To generate non-speech captions, we utilize pre-trained large language models (LLMs) to adapt NSI according to DHH viewers’ preferences. Yet LLMs often lack mechanisms for grounding, controllability, and user alignment. As a result, their outputs can diverge from creators’ intent or viewers’ goals \cite{noauthor_what_2021, shen_towards_2024, ji_ai_2024}—issues that are particularly consequential in accessibility contexts.

Prior work in human-AI interaction has explored various mechanisms for controlling generative models. Low-level techniques like temperature or top-k sampling adjust randomness but often lack semantic consistency \cite{holtzman2019curious, zhang2023survey}. Prompt engineering offers higher-level scaffolding \cite{reynolds2021prompt, wei2022chain}, but struggles to generalize across diverse user needs. More robust strategies include interactive controls—such as sliders, presets, and style grids—that allow DHH users to iteratively shape outputs \cite{ouyang2022training, lee2024design}. Other approaches apply constraints to guide generations within creator-defined bounds \cite{kumar2022gradient}. Post-generation methods such as rule-based filtering or semantic reranking also help enforce alignment \cite{gehman2020realtoxicityprompts, dinan2019build}, while human-in-the-loop feedback systems support refinement through ongoing user corrections \cite{amershi2014power, wang2021putting}. 

CapTune integrates several of these strategies by embedding creator- and viewer-defined parameters (e.g., expressiveness, level of detail, genre alignment) directly into the generative loop. This ensures that viewer preferences guide the output, while transformations remain within creator-defined boundaries.

\section{Understanding DHH Viewers' Needs and Preferences on Closed Captions of Non-Speech Information}

To understand how non-speech captioning practices align---or fail to align---with the lived experiences of DHH viewers, we conducted a qualitative content analysis of online discussions to understand the needs and preferences of DHH viewers regarding non-speech captions. Specifically, we aim to explore the following questions: (1) What closed caption-related factors influence DHH viewers' experience watching video content (e.g., movies, TV series, documentaries)? and (2) What are the actionable design opportunities to fulfill the identified needs and preferences?

\subsection{Data Collection}

Our data collection involved programmatically querying 12 subreddits (e.g., r/deaf, r/HardOfHearing, r/disability, r/accessibility, r/ClosedCaptioning, r/AskReddit) and six streaming service communities using PRAW \cite{noauthor_praw_2023}, an API for scraping Reddit content. We employed 11 search terms combining ``closed captions,'' and ``caption'' with DHH-related keywords such as ``DHH,'' ``deaf,'' and ``hard of hearing.'' This cross-sectional approach yielded 984 discussion threads (excluding duplicates). Two researchers audited these threads and removed off-topic content, including speech or subtitle-related concerns, non-DHH-related discussions, and posts about live captioning devices. Ultimately, our data comprised 51 posts from 13 unique threads, which contained first-person accounts of closed caption experiences from DHH viewers.

\subsection{Data Analysis}
We analyzed the filtered Reddit discussions following the open, axial, and selective coding process \cite{guest_applied_2012}. We examined each post as a coding unit and organized them into broader conceptual categories, identifying potential relationships between them. Two researchers met regularly to discuss and refine these emerging categories, resolving disagreements through consensus. This process led to 18 secondary concepts 
that captured the various dimensions of DHH viewers' preferences and experiences regarding non-speech captions (Appendix A). Finally, we integrated the secondary concepts into four distinct themes, which are presented in the following sections. Throughout this process, we paid particular attention to first-person accounts and specific examples shared by DHH viewers, as these provided authentic insights into their lived experiences with non-speech captions.

\subsection{Findings}
\label{sec:content-analysis-findings}
Our initial open coding identified 18 secondary concepts capturing the concerns, preferences, and experiences of DHH viewers. For example, the concept “Mood Conveyance” was illustrated by one viewer who noted: \textit{“Captions like ‘exciting music plays’ and ‘menacing laughs’ help me know the general mood of the scene. A lot of people don't realize that sound helps aid the general mood like danger or happiness.”} Similarly, “Sound Depiction Method” emerged from reflections like: \textit{“If it was just onomatopoeia for environmental sounds, like ‘bang bang,’ it would be hard for me to tell if it was from gunshots or a hammer or a cranky plumbing pipe.”} A full list of concepts appears in the Appendix.

Guided by these concepts, our analysis revealed several key insights into the factors shaping DHH viewers' experiences with non-speech captions. Quotes are drawn verbatim from participant interviews.

\subsubsection{Narrative and Emotional Engagement}
DHH viewers frequently described non-speech captions as critical to both narrative comprehension and emotional engagement with content. Eight viewers specifically highlighted the importance of music captions in shaping their understanding of a scene's emotional tone. One participant noted: \textit{``A scene is very different if it's happy music versus suspense music. If it's the psycho music you know someone's about to die versus happy music... where no one is going to die.''} This supports prior work suggesting that well-captioned sound information can enrich storytelling for DHH audiences \cite{vy_using_2010}. Beyond music, seven viewers emphasized the value of genre-specific cues. For example, when asked about how captions like ``menacing laugh'' aided their understanding of the content, P2 answered: ``\textit{If you are not able to hear the mood music change for a scare, then it half scares you.}''

DHH viewers also valued paralinguistic information for conveying emotional tone. For example, one user listed an example of the ideal caption: ``[John, sarcastically] Oh yes, definitely that, let's do that.'' and stated: ``\textit{Without the sarcasm tone marker, that sentence is completely different in meaning.}'' Another viewer pointed out that this is especially important if the tone is ``not obvious from what they see on-screen.'' This aligns with \citeauthor{rashid_dancing_2008}'s work, which demonstrates the importance of paralinguistic information in helping DHH viewers understand speaker intentions and emotional states \cite{rashid_dancing_2008}.



\subsubsection{Sound Perception and Representation}
DHH viewers reflected on their perception of sound information and their preferences on how sound information should be presented in closed captions, often shaped by their diverse personal backgrounds and hearing history. For example, a user stated: ``\textit{A person I know is profoundly deaf and is amused by things like [dramatic music]. As one said to me once: How would I know what dramatic music is, and why would I care?}'' These reflections echoed prior work demonstrating the individualized nature of caption preferences \cite{udo_rogue_2010} and, more broadly, that users' preferences for assistive technologies are heavily influenced by individual lived experiences \cite{shinohara_shadow_2011}.

In terms of preferences for sound representation methods, Onomatopoeic descriptions (e.g., ``Thud'', ``Swoosh'') were preferred by four viewers, while descriptive text focusing on sound sources or sensory qualities was favored by seven participants; one of these viewers stated: 
\begin{quote}
    \textit{If it was just environmental sound ``bang bang'' it would be hard for me to tell if it's from gunshots or a hammer or a cranky plumbing pipe... So a description is more effective, like [gunshots] and [pipe banging].}
\end{quote}

\subsubsection{Information Density and Contextual Clarity}
DHH viewers have expressed diverse preferences regarding what and how much non-speech sound information should be included in closed captions. For example, one viewer responded to an inquiry about \textit{``pet peeves''} in movie captions: ``\textit{Do not use vague descriptions (or none  at all) of the background noise,}'' while another viewer found overly captions with detailed sound descriptions \textit{``distracting to look at.''}

Importantly, DHH viewers' considerations about the density of sound information in closed captions extended beyond simple quantity considerations to encompass plot and contextual relevance, cognitive demands (e.g., reading speed and attention split from the scene), and redundancy avoidance (e.g., displaying information that exists in the visual scene). Plot and contextual relevance emerged as primary criteria for determining the appropriate information density in closed captions. For example, one viewer noted: ``\textit{I hate it when captioning misses important auditory information like `knock on door,' or anything that's important to the story.}'' Indeed, DHH viewers rely heavily on captions to fill gaps in narrative comprehension that hearing viewers obtain through incidental auditory information \cite{cambra_how_2010}.

In terms of cognitive processing, some viewers expressed concerns about how caption reading may compete with visual scene processing. For example, one viewer noted:
\begin{quote}
    \textit{If I can't hear the music, describing it doesn't add anything and actually begins to break my immersion when, for example, the scene has two people staring at each other intensely but no dialogue, and my eyes keep constantly flicking down because there's a caption but it's just describing the music.}
\end{quote}

Some viewers also cautioned that captions should not duplicate information that is already visually available. As one participant explained: ``\textit{DON'T say: `bad guys firing at the car.' Say [rapid gunfire], [intermittent gunshots].}'' Another echoed this sentiment, noting that describing tones of in-screen speakers or facial expressions is unnecessary as they can ``\textit{see if they are being rude or friendly.}''

\subsubsection{Stylistic Preferences}
Beyond informational content, many DHH viewers had strong preferences regarding the style of non-speech captions. Many appreciated creative, evocative captions; for example, one viewer used \textit{Stranger Things}, a popular Netflix series, as an exemplary approach to non-speech captioning: \textit{“‘Industrial synth music hums with lots of squelching’— [they] took this to a new level.”} Another shared: \textit{``I watched a show last night where the captioner really got into it: `sinuous music,' `mysterious music,' `lively happy music'... they better give that captioner a raise.''} On the other hand, one viewer found these expressive captions \textit{``distracting.''} Despite the diversity, for some, stylistic preference are extended beyond entertainment---it reflected deeper desires for cultural participation and emotional resonance, which aligns with the idea of \textit{access intimacy}, the experience of having one's accessibility needs met in a way that feels intuitive and natural \cite{noauthor_access_2011}.

\subsection{Summary and Actionable Design Opportunities}
Our analysis reveals that DHH viewers have diverse preferences for non-speech captions, influenced by their background, hearing history, and individual needs. This suggests that static captioning approaches are insufficient. We distill our findings into four key design opportunities for customizable captions:

\begin{enumerate}
    \item \textbf{Varying Levels of Detail:} While detailed captions support comprehension, overly verbose descriptions can disrupt immersion or tax cognitive load. Adjustable detail levels enable viewers to tune caption density based on personal preference and narrative relevance.

    \item \textbf{Expressiveness for Emotional Engagement:} Captions were viewed not just as informative, but as expressive tools conveying tone and emotion. Cues like \texttt{[sinuous music]} or \texttt{[John, sarcastically]} helped viewers interpret mood. Some, however, found them distracting, underscoring the need for customization.

    \item \textbf{Personalizable Sound Representation:} Viewers had distinct preferences—some favored onomatopoeia, others sensory descriptors, or source-based labeling. These stemmed from varied familiarity with sound metaphors, underscoring the need for multiple representation modes.

    \item \textbf{Genre Alignment for Thematic Fit:} Stylized captions helped convey genre tone (e.g., “menacing laugh” in horror), while generic ones undermined narrative coherence. Allowing viewers to opt into genre-aligned captioning helps maintain thematic consistency.
\end{enumerate}

These four themes directly informed the four customization parameters implemented in our prototype later (Section 4.1). Each reflects a distinct axis along which DHH viewers' preferences vary, as surfaced through their lived experiences shared in our analysis.

\section{The CapTune System}
We present CapTune, a system that allows DHH viewers to personalize non-speech captions while preserving creators' intent. Rather than relying on predefined rules or static templates, CapTune leverages large language models guided by creators' specifications and viewer preferences to dynamically adapt captions, preserving narrative coherence and creative intent while addressing diverse accessibility needs.

CapTune comprises two primary components:
\begin{itemize}
    \item \textbf{Creator Tool (CT): } Allows creators to define acceptable transformation boundaries for non-speech captions.
    \item \textbf{Viewer Client (VC): } Enables DHH viewers to personalize captions within the creator-defined boundaries.
\end{itemize}

To illustrate the system in action, we follow Sam, a creator working on captions for \emph{Bella}, a short animated film about a stray cat's journey toward adoption, and Jamie, a DHH viewer who experiences the video using the customized captions.

\begin{figure*}[ht]
    \centering
    \includegraphics[width=\linewidth]{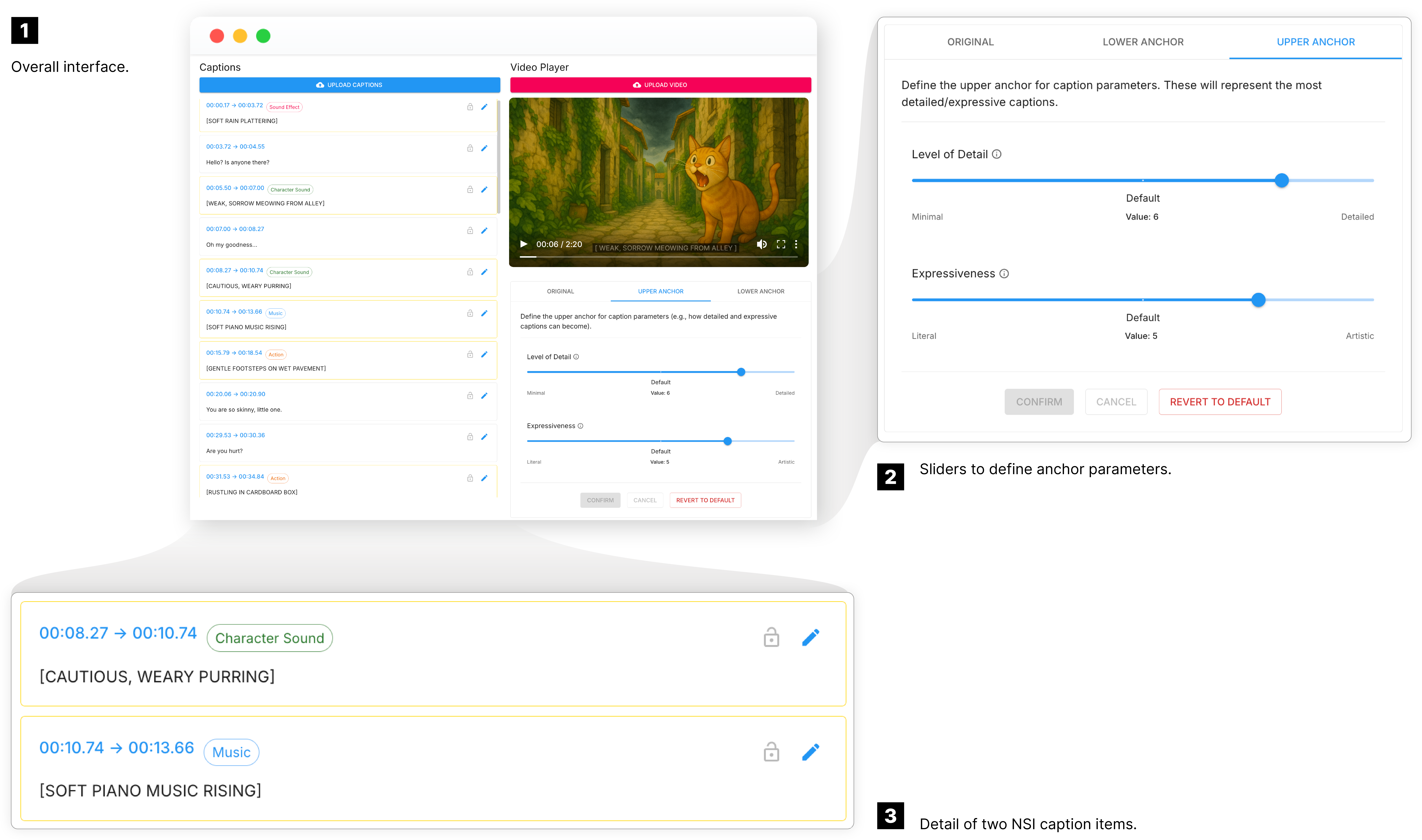}
    \caption{Creator Tool interface for configuring the caption transformation space. The image shows (1) the overall interface with a caption list, video player, and anchor configuration panel; (2) sliders for setting anchor values for level of detail and expressiveness; and (3) a detailed view of two non-speech information caption items.}
    \Description{Screenshot of the Creator Tool interface used to configure the transformation space for caption customization. Label 1 marks the full interface, which includes a video player, a caption list, and a panel for adjusting anchor parameters. The caption visible in the video is “[WEAK, SORROW MEOWING FROM ALLEY].” Label 2 highlights sliders for setting the upper anchor values for Level of Detail and Expressiveness, both set near the middle of the scale. Label 3 zooms into two non-speech caption items in the list: “[CAUTIOUS, WEARY PURRING]” labeled as “Character Sound,” and “[SOFT PIANO MUSIC RISING]” labeled as “Music.”}
    \label{fig:creator_tool_overview}
\end{figure*}

\subsection{Caption Customization Parameters}
Informed by our analysis of DHH viewers' caption preferences (Section~\ref{sec:content-analysis-findings}), CapTune supports four key customization parameters:
\begin{enumerate}
    \item \textbf{Level of Detail: } Adjusts information density in captions on a scale of 1 to 10. Lower values provide concise descriptions (e.g., \texttt{[Music playing]}), whereas higher values deliver rich, detailed descriptions (e.g., \texttt{[Soft piano melody with rising tempo]}).
    \item \textbf{Expressiveness: } Adjusts stylistic language on a scale of 1 to 10. Lower values yield neutral, straightforward captions (e.g., \texttt{[Door closes]}); higher values generate evocative, artistically expressive captions (e.g., \texttt{[Door slams shut with a resonant thud]}).
    \item \textbf{Genre Alignment: } Optional parameter enabling captions to stylistically align with a video's genre or narrative mood (e.g., transforming \texttt{[Wind blowing]} into \texttt{[Whimsical swoosh]} or \texttt{[Freezing wind]} in an animated fantasy).
    \item \textbf{Sound Representation: } Offers three distinct methods for depicting sounds:
    \begin{itemize}
        \item \textbf{Source-focused: } Emphasizes the sound source.
        \item \textbf{Onomatopoeia: } Uses phonetic imitations of sounds.
        \item \textbf{Sensory quality-focused}: Emphasize sensory characteristics like pitch, texture, or intensity.
    \end{itemize}
\end{enumerate}

The first two parameters---\textsc{Level of Detail} and \textsc{Expressiveness}---are controlled by creators to establish the ``safe'' range of transformations, as these dimensions significantly influence narrative tone and content density. \textsc{Genre Alignment} and \textsc{Sound Representation} are fully viewer-controlled, as these stylistic choices do not risk altering the intended narrative meaning. Examples of how each parameter can modulate textual descriptions of sounds are shown on Figure~\ref{fig:teaser}.


\subsection{Creator Tool: Defining Transformation Space for Non-Speech Captions}
The Creator Tool (Figure~\ref{fig:creator_tool_overview}) enables creators to specify how much and in what ways their captions can be transformed. Rather than authoring all possible variations, creators define a two-dimensional space---bounded by parameters of \textsc{Level of Detail} and \textsc{Expressiveness}---within which AI-transformations can occur.

\subsubsection{Setup and Context Extraction}
When Sam uploads the animation and its caption file (in \texttt{.srt} format), the system displays the captions in a structured list view. Each entry includes timestamps, NSI category tags (e.g., ``character sounds,'' ``music''), and caption text. The system automatically detects non-speech caption---e.g., those enclosed in brackets or parentheses---and visually highlights them with yellow borders. CapTune then extracts relevant audio-visual context for each highlighted caption. Specifically, it generates a video segment starting 5 seconds before and extending 5 seconds after each caption's timestamp. These segments are then processed by VideoLLaMA2, a video large language model capable of decoding and understanding combined audio-visual information, with the prompt: ``Can you describe the visual and audio content in this video clip?'' VideoLLaMA2 outputs detailed textual descriptions of each segment's visual and audio content, providing essential context for accurate caption transformations.



\begin{figure*}[ht]
    \centering
    \includegraphics[width=\linewidth]{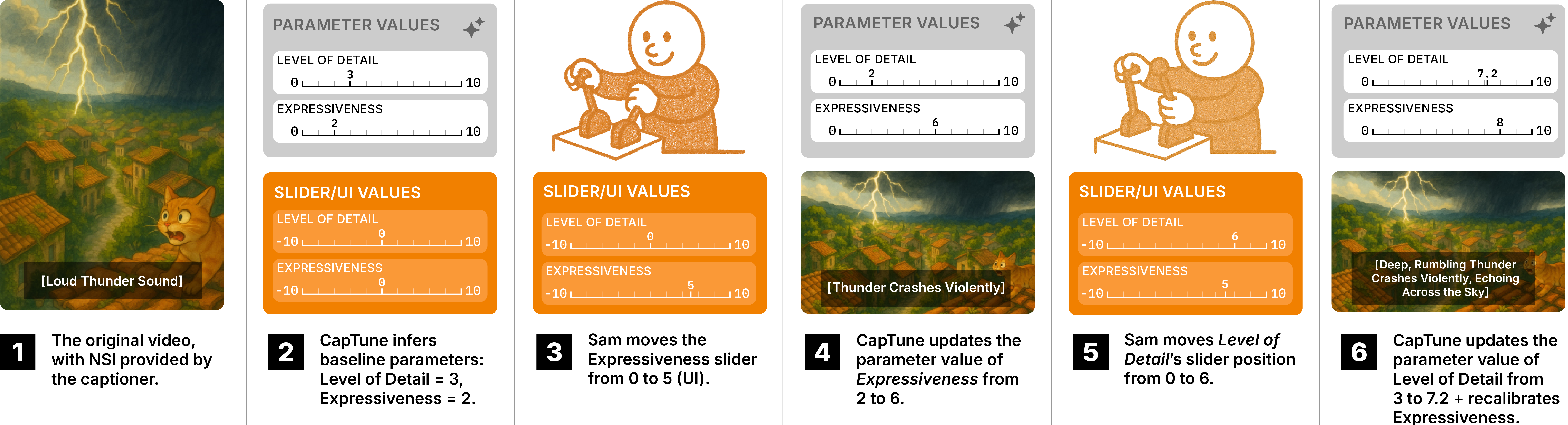}
    \caption{Example of iterative caption transformation in CapTune. (1) The original caption, [Loud Thunder Sound], is provided by the creator. (2) The system analyzes this caption and infers initial parameter values of \textsc{Level of Detail = 3} and \textsc{Expressiveness = 2}. (3) Sam moves the slider position for \textsc{Expressiveness} to 5. (4) The system updates \textsc{Expressiveness}'s parameter value to 6, generating the transformed caption: [Thunder Crashes Violently]. (5) Sam now moves \textsc{Level of Detail}'s slider position from 0 to 6. (6) The system updates the parameter value of \textsc{Level of Detail} to 7.2; the caption now becomes [Deep, Rumbling Thunder Crashes Violently, Echoing Across the Sky]. GPT-4o recalibrates \textsc{Expressiveness}'s parameter value to 8 and maps this value to its current slider position, 5, to maintain UI consistency.}
    \Description{A step-by-step visual diagram showing how a non-speech caption is iteratively transformed as part of the anchor creation process in the Creator Tool. Step 1 shows a stormy scene with the original caption “[Loud Thunder Sound].” Step 2 displays the system’s inferred internal values for Level of Detail (3) and Expressiveness (2). Step 3 introduces the viewer interface, where Sam increases Expressiveness to 5, which maps to 6 internally. Step 4 shows the caption updated to “[Thunder Crashes Violently],” and this new state becomes the recalibrated baseline. In Step 5, the viewer increases the Level of Detail to 6. Step 6 shows the caption updated to “[Deep, rumbling thunder crashes violently, echoing across the sky]” and the system re-evaluating Level of Detail as 7.2 and Expressiveness to 8. A small cartoon figure with orange outlines appears in Steps 3 and 5, indicating viewer interaction with the caption style sliders.}
    \label{fig:model}
\end{figure*}

\subsubsection{Setting Transformation Baselines}
Next, CapTune analyzes Sam's original captions using GPT-4o to determine baseline values for \textsc{Level of Detail} and \textsc{Expressiveness} on a scale of 1 to 10. For the \textit{Bella} animation, the system determines that Sam's original captions have a moderately low amount of details and neutral language, assigning the baseline values as \textsc{(Level of Detail = 3, Expressiveness = 2)}. These values serve as a calibration point---the ``original'' state from which all subsequent transformations are measured.

\subsubsection{Defining Transformation Space with Anchor Points}
\begin{figure*}[ht]
    \centering
    \includegraphics[width=0.95\linewidth]{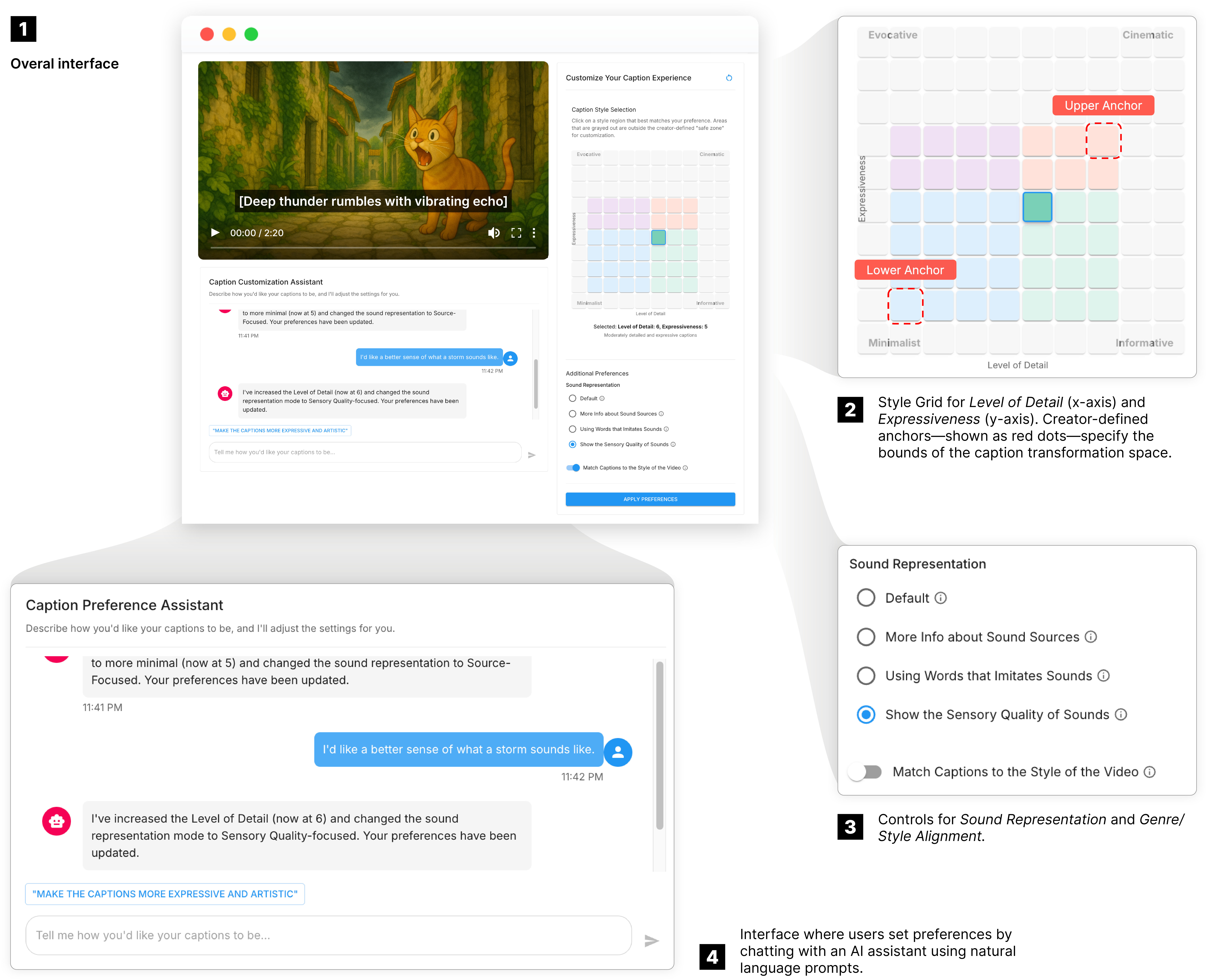}
    \caption{Viewer Client interface for customizing non-speech information in captions. The image shows (1) a screenshot of the overall interface, highlighting (2) a grid where users define their preferred captioning style along two dimensions—\textsc{Level of Detail} and \textsc{Expressiveness}. Available options are constrained by the creator and shown as colored cells within the grid. Also shown is (3) a panel where users choose how sound events are represented in text, including a default setting, more information about sound sources, onomatopoeia, and sensory qualities of sound. A toggle allows captions to match the genre or style of the video. Lastly, (4) a chat interface enables interaction with an AI assistant that helps to understand and adjust settings through natural language prompts.}
    \label{fig:zoom_viewer}
    \Description{Screenshot of the Viewer interface for customizing non-speech captions. Label 1 marks the overall interface, which includes a video player, a style grid, sound representation options, and a chat interface with an AI assistant. The video player shows the caption "Deep thunder rumbles with vibrating echo." Label 2 points to a style grid for adjusting Level of Detail (x-axis) and Expressiveness (y-axis), with color-coded cells and quadrant labels such as Minimalist, Evocative, Informative, and Cinematic. Label 3 highlights controls for choosing sound representation styles, including default, source-based, onomatopoeic, sensory-focused, and genre-aligned options. Label 4 identifies the interface where users set preferences by chatting with an AI assistant using natural language prompts. The example prompt reads “I’d like a better sense of what a storm sounds like,” and the assistant responds by saying “I’ve increased the Level of Detail (now at 6) and changed the sound representation mode to Sensory Quality-focused. Your preferences have been updated.”}
\end{figure*}

Sam defines the transformation space by setting two anchor points that represent the acceptable range for viewer customization. The lower anchor sets minimum \textsc{Level of Detail} and \textsc{Expressiveness}, representing the most minimal captions that can still maintain clarity, while the \textbf{upper anchor} marks the most elaborate and expressive version Sam is comfortable with. These anchors delineate a two-dimensional space within which all transformations must remain (visualized in Figure~\ref{fig:zoom_viewer}.2).

Sam adjusts each anchor using a slider interface. Slider movements trigger real-time caption previews generated by GPT-4o, allowing Sam to assess the impact of the parameter changes. For example, starting from an original caption, \texttt{[Loud thunder sound]}, Sam increases the \textsc{Expressiveness} slider from 0 to 5, producing a transformed caption: \texttt{[Thunder crashes violently]}. Sam then increases the \textsc{Level of Detail} slider from 0 to 6, producing the caption: \texttt{[Deep, rumbling thunder crashes violently, echoing across the sky]}.

Behind the scenes, CapTune decouples the visual slider interface from the model's underlying parameter values. While the sliders operate in a symmetrical UI range (from -10 to 10), the internal parameter values span from a semantic scale of 1 to 10 for both \textsc{Level of Detail} and \textsc{Expressiveness}. When Sam adjusts the slider position, the system uses a piecewise linear mapping function that translates slider positions ($s$) to new parameter values ($f(s)$):
\begin{equation}
f(s) = 
\begin{cases}
\begin{split}
&V_{\text{min}} + \frac{s - s_{\text{min}}}{s_0 - s_{\text{min}}} \times (V_0 - V_{\text{min}})
\end{split} & \text{if } s < s_0 \\[0.5em]
\\
\begin{split}
&V_0 + \frac{s - s_0}{s_{\text{max}} - s_0} \times (V_{\text{max}} - V_0)
\end{split} & \text{if } s > s_0
\end{cases}
\end{equation}
where $s_0$ is the center slider position (set to 0) representing the original caption's baseline parameter value, $V_0$. The ranges $[s_{\text{min}}, s_{\text{max}}]$ and $[V_{\text{min}}, V_{\text{max}}]$ define the bounds for slider positions and parameter values, respectively. In our implementation, we set $[s_{\text{min}}, s_{\text{max}}] = [-10,10]$ and $[V_{\text{min}}, V_{\text{max}}] = [1,10]$, though these values can be adjusted for different future applications. This mapping ensures proportional scaling while maintaining intuitive slider behavior.

Notably, this system acknowledges potential interaction effects between \textsc{Level of Detail} and \textsc{Expressiveness}---two parameters that, although conceptually distinct, may influence each other in practice. For instance, increasing detail in a caption may inadvertently make it sound more expressive. To address this, the system employs a recalibration strategy: when Sam adjusts \textsc{Level of Detail}, GPT-4o recalculates the parameter value for \textsc{Expressiveness} based on the transformed captions. This new value is then mapped to the current slider position for \textsc{Expressiveness}. This approach preserves visual consistency on the interface level while acknowledging semantic shifts. Figure~\ref{fig:model} visualizes this recalibration process.

Sam can continue adjusting the sliders to explore a range of caption variations. If a generated transformation is close but not quite right, Sam can manually edit it to achieve the desired result. Sam can also ``lock'' individual caption items, preventing them from being altered in future transformations. Together, these mechanisms allow creators to maintain fine-grained editorial control while leveraging the efficiency of AI assistance.

\subsubsection{Export and Distribution}
Once the anchors are finalized, Sam exports a configuration file in JSON format. This includes: the original captions, lower and upper anchors for \textsc{Level of Detail} and \textsc{Expressiveness}, parameter values for each, audio-visual descriptions for video segments, and the video metadata (e.g., genre, title, and brief descriptions of \textit{Bella}). This file becomes the basis for real-time caption transformation in the Viewer Client.

\subsection{Viewer Client: Personalizing Non-Speech Caption Experiences}

The Viewer Client (VC) enables DHH viewers like Jamie to personalize non-speech captions in real-time based on their preferences for \textsc{Level of Detail}, \textsc{Expressiveness}, \textsc{Genre Alignment}, and \textsc{Sound Representation} (Figure~\ref{fig:zoom_viewer}). Viewers can specify their preferences through direct manipulation or natural language interactions. Transformations occur seamlessly during video playback.


\subsubsection{Direct Manipulation Interface}

The direct manipulation interface features a 10\,$\times$\,10 style grid representing the creator-defined transformation space, with \textsc{Level of Detail} on the horizontal axis and \textsc{Expressiveness} on the vertical axis. The lower and upper anchor points are represented by the lower-left corner and the upper-right corner of the colored fields (as indicated in Figure~\ref{fig:zoom_viewer}.2), respectively. The grid provides reference points at the corners: Minimalist (low detail, low expressiveness), Informative (high detail, low expressiveness), Evocative (low detail, high expressiveness), and Cinematic (high detail, high expressiveness), though most selections fall somewhere between these extremes. Cells outside the creator-defined boundaries are disabled, ensuring transformations stay within acceptable limits. In Jamie's case, the available space spans from 2 to 8 for the \textsc{Level of Detail} and 2 to 7 for \textsc{Expressiveness}. By default, the cell corresponding to the original caption's parameter values (\textsc{Level of Detail = 3, Expressiveness = 2}) is selected. By selecting the cell (\textsc{Level of Detail = 6, Expressiveness = 5}), Jamie transforms \texttt{[Loud thunder sound]} to \texttt{[Rumbling thunder crashes violently]}. If Jamie instead selects a cell like (\textsc{Level of Detail = 1, Expressiveness = 2}), the caption becomes \texttt{[Thunder]}. 

In addition to the style grid, Jamie can further tailor the experience by toggling \textsc{Genre Alignment} and \textsc{Sound Representation}. Turning on \textsc{Genre Alignment} adapts the captions to match \textit{Bella's} animated style---so a caption like \texttt{[Loud thunder sound]} might become \texttt{[Scary thunder crashes]}. Since Jamie wants to experience the sound in a more visceral way, he switches the \textsc{Sound Representation} to ``Onomatopoeia,'' changing the caption to \texttt{[BOOM! Thunder crashes]}.

\subsubsection{Natural Language Interface}
Alternatively, Jamie can specify preferences using conversational input. For example, Jamie tells the system: ``I want to know what is making the sounds, but keep it brief.'' The system interprets as a request for lower \textsc{Level of Detail} and source-based representation and updates the parameters accordingly. Jamie might then see \texttt{[Door creaks]} instead of \texttt{[Continuous creaking sound]}. Later, during a storm scene, Jamie types: ``I'd like a better sense of what the storm sounds like.'' The system then raises \textsc{Level of Detail} and selects the ``sensory-focused'' representation option, transforming \texttt{[Wind blowing]} to \texttt{[Wind whistles with chilling howl]}.

\subsubsection{Caption Transformation Process}
The system leverages GPT-4o to transform captions based on Jamie's specifications. To achieve reliable transformations, the system quantifies the proposed changes by calculating interpolation ratios $r$ that measure where the new parameter value ($V'$) falls between the lower ($V_{\text{min}}$) and upper anchors ($V_{\text{max}}$):
\begin{equation}
    r = \dfrac{V' - V_{\text{min}}}{V_{\text{max}} - V_{\text{min}}} \\
    \quad \text{where} \; r \in [0,1]
\end{equation}
This determines the relative positioning of the request within the creator-defined transformation space.

Additionally, the system computes the normalized change ratio $\delta$ that measures the degree of change from the current setting ($V$) to the new preference ($V'$):
\begin{equation}
    \delta = \dfrac{V' - V}{V_{\text{max}} - V_{\text{min}}} \\
    \quad \text{where} \; \delta \in [-1,1]
\end{equation}
This ratio reflects both the magnitude and direction of change from the current caption.

These ratios are computed independently for both \textsc{Level of Detail} and \textsc{Expressiveness}. When Jamie moves the selection from \textsc{(Level of Detail = 3, Expressiveness = 2)} to \textsc{(Level of Detail = 6, Expressiveness = 5)}, this new preference result in $(r=0.67, \delta=0.50)$ for \textsc{Level of Detail}, and $(r=0.60, \delta=0.60)$ for \textsc{Expressiveness}. These values are then used to construct a prompt that explicitly quantifies the intended transformation:

\begin{small}
\begin{verbatim}
Please transform the current caption based on the following 
specifications:
  -- Level of Detail
        67% more detailed than the [lower-anchor captions]
        33% less detailed than the [upper-anchor captions]
        50% more detailed than the [current captions]
  -- Expressiveness
        60% more expressive than the [lower-anchor captions]
        40% less expressive than the [upper-anchor captions]
        60% more expressive than the [current captions]
\end{verbatim}
\end{small}

The system then passes this prompt to GPT-4o along with the original caption, lower- and upper-anchor captions, and relevant scene context from the video. This structured prompt enables the model to interpolate between concrete reference points rather than generating captions from scratch, preserving both viewer intent and creator-defined constraints.

\subsection{Implementation Details}
The CapTune system, including both the Creator Tool and Viewer Client, was deployed on a server running Windows 11 Enterprise (Intel i5, 14-core CPU, NVIDIA GeForce RTX 4090 GPU). We built the front-end interface using React and TypeScript. The back-end included a Node.js server and a Python module. The Node.js server handled decoding caption files, parsing user preferences, and communicating with the GPT-4o model (via the OpenAI API) for caption transformations and natural language interpretations. The Python module analyzed audio-visual content of segmented video clips by communicating with a VideoLLaMA 2 model (VideoLLaMA2.1-7B-AV) hosted on a remote server. The output, which contained a textual description of the analyzed scenes, would then be transmitted to the Node.js server as JSON files. For more details, the full system codebase is open-sourced\footnote{GitHub.com/SoundabilityLab/CapTune}.

\section{Creator Tool Evaluation}
We evaluated the Creator Tool with participants who have experience creating and editing videos and captions. Our goal was to assess the usability of the system interface and explore creators' perceptions of AI-mediated caption creation and transformation.

\subsection{Method}
\subsubsection{Participants}
Upon obtaining IRB clearance, we recruited seven participants with diverse experiences in creating videos across various genres, including independent films, advertisements, travel and lifestyle blogs, tech reviews, music videos, and educational videos such as tutorials and technology demonstrations. These participants had an average of 6.4 years of experience (SD=3.4) in creating videos. The detailed demographics are listed in the Appendix (Table~\ref{tab:creator-demographics}).

\subsubsection{Procedure}
All sessions were conducted remotely via Zoom and lasted approximately 60 minutes. Each began with a 5-minute overview of transformation space concepts and a 5-minute walkthrough of the Creator Tool interface, including key workflows such as adjusting anchor points and exporting configuration files. Participants were then provided with three video clips—\textit{Kitbull} (an 8-minute Pixar animation), a 3-minute dolphin documentary, and a 3-minute clip from Frozen. These clips were selected to represent a diverse range of genres, sound characteristics, and viewing contexts. For each clip, participants completed three guided tasks: (1) defining a lower anchor point representing the minimal acceptable caption version, (2) defining an upper anchor representing the richest acceptable version, and (3) exporting the configuration file with accompanying metadata (e.g., title, genre, and brief storyline description).

We used a think-aloud protocol \cite{baxter_chapter_2015} during task completion. Afterward, participants took part in semi-structured interviews that covered their overall impressions, usability concerns, conceptual understanding of the transformation space, perceptions of creative control, factors influencing anchor creation, and considerations for real-world integration. All sessions were recorded and transcribed.

\subsection{Analysis}
Our data included video recordings and audio transcripts, which we analyzed using open, axial, and selective coding \cite{guest_applied_2012}. We began with open coding to identify meaningful units related to participants’ feedback and system interactions. This process yielded 72 distinct open codes, capturing observations such as ``difficulty interpreting effects after parameter adjustments,'' ``appreciated manual editing capabilities,'' and ``desired localized parameter controls for different scenes.'' We then applied axial coding to group related codes and uncover broader conceptual categories. Through iterative refinement and team discussion, we consolidated the open codes into eight second-level themes reflecting key aspects of creators’ experiences, including ``UI feedback,'' ``creative control and agency,'' and ``context-dependent preferences and adaptability.'' These themes informed a comprehensive understanding of how creators engaged with the Creator Tool.

\subsection{Findings}
Creators' experiences shed light on how the tool enabled meaningful control over AI-assisted caption transformations, while also surfacing key areas for improvement related to feedback clarity, parameter tuning, and integration into existing captioning practices.

\subsubsection{Conceptual Understanding of the Caption Transformation Model}
All participants easily grasped the metaphor of using upper and lower anchors to guide AI-generated transformations. While C2 initially proposed defining four anchors to better reflect a “2D transformation space,” his opinion evolved near the end of the session: \textit{``Yeah, two is probably enough. I don't think I will need to do another two.''}

Creators recognized that the ideal parameter settings varied by content type (N=5), scene pacing (N=3), and communication goals (N=1). C3 explained that for emotionally-driven narratives, \textit{``expressiveness plays a very important role,''} but for educational content, he preferred to \textit{``tune it down a bit''} to maintain clarity. C6, reviewing a scene with fast-paced sound effects with \textsc{(Level of Detail = 7, Expressiveness = 8)}, said, \textit{``I mean, even I could not finish that... I imagine it will be painful to read through the captions in this scene.''} C6 recommended adapting captions' information density based on scene complexity. Four participants also suggested moving beyond global anchor settings to allow for scene-specific configurations. As C4 put it, \textit{``There may be cases where I want a less detailed caption even though you might want the entire clip to be expressive.''}

\subsubsection{Creative Control and Semantic Integrity}
All participants emphasized the importance of retaining creative agency when using AI to transform captions and appreciated how the system supported this. C2 described CT's workflow as a way to \textit{``balance accessibility and creative control.''} C1 echoed this sentiment, noting that because the system's anchor-setting process was \textit{``not fully automatic,''} it allowed creators to remain actively involved in shaping the outputs. C6 further reinforced this perspective, pointing to the manual edit-and-lock feature as a meaningful safeguard: it enabled her to ensure that specific phrasings aligned with the intended narrative, rather than relying solely on AI interpretations.

When interacting with CT, five creators were particularly attentive to the semantic accuracy in caption transformations. C2 flagged an output that transformed \texttt{[Anna exhales]} to \texttt{[Anna exhales with a sigh of relief]}, noting this interpretation was \textit{``overly specific''} and might \textit{``flatten the scene's emotional ambiguity.''} C6 similarly questioned the change from \texttt{[Ice freezing sound]} to \texttt{[Chill ice cracking]}, noting: \textit{``That might mislead people... It's not really a `cracking' sound.''} These concerns reflect creators' commitment to preserving narrative nuance.

\subsubsection{System Utility and Workflow Integration}
\label{sec:ct-utility}
Participants found the tool intuitive (N=5), with a gentle learning curve (N=2), and saw potential for improving caption accessibility (N=3). C2 noted its labor-saving value: \textit{``There is no way I'm gonna sit and plot through like 400 different captions,''} adding, \textit{``this workflow is a lot faster. And if AI generation becomes more reliable, I can imagine it being automated completely.''} C7 framed the act of creating anchors as an \textit{``ethical responsibility''} tied to accessibility and emphasized that seamless integration with professional tools---like Davinci Resolve---would be key to adoption.


Despite the intuitive design, participants wanted clearer feedback on transformation outcomes. C4 found it difficult to differentiate the effects between minor adjustments (e.g., changing \textsc{Expressiveness} from 6 to 7) and requested a more concrete indication of change. C7, after lowering \textsc{Level of Detail} from 0 to -6, remarked: \textit{``Since it's already very short, I am not noticing a lot of changes.''} C2 reported a similar concern and proposed shrinking the slider range to $[-5,5]$ and called for \textit{``more transparent communication of the transformation logic''} to better ``debug'' \cite{he_err_2024} the outputs.

Participants also suggested several workflow extensions. For example, C6 envisioned a prompt-based audit tool (e.g., \textit{``Can you check and make sure captions for these themes do not use `flowery' descriptions?''}) C2 observed that some scenes had two related non-speech captions---e.g., \texttt{[Thunder roars]} and \texttt{[Dramatic music]}---and proposed a mechanism for merging them into a unified caption (e.g., \texttt{[Thunder crashes amid dramatic music]}).



\section{DHH Viewer Client Evaluation}
We evaluated the Viewer Client with 12 DHH participants to explore their views on AI-customized non-speech captions.

\subsection{Participants}
We recruited 12 DHH participants (seven female, five male; aged 23–67, M=38.3, SD=16.7) who regularly use closed captions when consuming video content. Participants self-identified across a range of identities, including six as Deaf, four as Hard of Hearing, and two as deaf. Most reported profound hearing loss (N=8), with others describing their hearing loss as severe (N=2) or asymmetric (e.g., severe in one ear, none in the other; N=1). Participants also reported diverse communication backgrounds: while most used English as their primary language (N=9), three reported using ASL as a primary or secondary language. To assess baseline satisfaction with current non-speech captions in mainstream media, participants rated their experiences on a Likert scale. Three participants were ``not at all satisfied,'' six were ``slightly dissatisfied,'' and two were ``moderately satisfied.'' The detailed demographics are listed in the Appendix (Table~\ref{tab:viewer-demographics}).

\subsection{Procedure}
The IRB-approved study lasted approximately 90 minutes and was conducted remotely via Zoom. Sign language interpreters and CART captioners were available as requested. Sessions were also video recorded with participant consent and later transcribed for analysis.

The session began with an introduction to the research goals and logistics, as well as the consent procedures. Participants were then introduced to the CapTune Viewer through a brief demonstration that explained the four transformation parameters and interface controls. We took care to explain concepts that might be unfamiliar, such as ``onomatopoeia'' and ``sensory qualities of sounds.'' The core of the session involved hands-on exploration with the system. Participants were encouraged to interact with the VC while watching the same video clips we provided during the creator evaluation (i.e., \textit{Kitbull}, the dolphin documentary, and a clip from \textit{Frozen}). During the exploration, we employed a think-aloud protocol \cite{baxter_chapter_2015}, asking participants to verbalize their thoughts and reactions as they experimented with different settings. Participants fully explored the system by selecting different grid cells and toggling \textsc{Sound Representation} options and \textsc{Genre Alignment}.

After using the system, participants took part in a semi-structured interview that covered their overall impressions, preferences regarding the four customization parameters across different content types, and their perceived utility and feasibility of the CapTune pipeline. We also explored potential use cases where participants felt the system would be most beneficial, and gathered suggestions for improving the interface and workflow. Participants were compensated \$40/hour for their time.

\subsection{Data Analysis}
All sessions were transcribed verbatim, with ASL interpretations translated into English. We analyzed the transcripts following \citet{guest_applied_2012}'s Applied Thematic Analysis approach. To begin, the first author skimmed the transcripts to become familiar with the data and collaborated with the research team to develop an initial codebook. The researcher then applied and refined the codes through an iterative coding process. The finalized codebook consisted of a three-level hierarchy, comprising 10 first-level, 22 second-level, and 168 third-level codes. A second researcher independently applied the final codebook to all transcripts. Interrater reliability (IRR) measured using Cohen's Kappa \cite{mchugh_interrater_2012} was, on average, $0.74$, and the raw agreement was $90.5\%$. Discrepancies were resolved through consensus.

\subsection{Findings}

Our analysis of 12 DHH participants' experiences with the Viewer Client revealed several key insights into system usability, customization preferences, and envisioned use cases.

\subsubsection{Viewer Impressions}
\label{sec:viewer-ux}
Participants generally responded positively to the Viewer Client, feeling that the customization deepened their emotional connection to the content. Most participants (N=9) felt the system enhanced their engagement. P7 stated: \textit{``I was watching the video, and the captions actually gave me goosebumps. It really did help.''} Reflecting on music captions, she added: 
\begin{quote}
    \textit{It's so annoying when captions from other sources just say song or music. So when they say it's `tense music,' that gives a lot more feeling to the song. It's definitely much more emphatic.}
\end{quote}

The chat interface was also well-received, though used less frequently. P7 described it as \textit{``additional support''} that helped her better understand the system's capabilities. Participants' queries, such as a playful request to \textit{``describe the sounds like Shakespeare,''} revealed opportunities for future expansion. However, only three participants used the chat interface, suggesting that while it was appreciated as a supplementary feature, most viewers gravitated toward the direct visual controls.

\subsubsection{Usability Challenges and Desired Enhancements}
While the core ideas were praised, participants identified several usability challenges. Seven participants found the process of tuning parameters to be a matter of trial-and-error that \textit{``took the attention away from the video itself''} (P5).

Participants also offered feedback on the granularity of the controls. Six participants found the $10\times10$ style grid overly dense or difficult to navigate. P5 remarked, \textit{``The scale of 1 to 10 is a bit too much,''} noting that subtle parameter changes (e.g., adjusting \textsc{Level of Detail} from 3 to 4) were not always perceptible. Four participants stated it was difficult to understand \textit{``what exactly changed''} after making adjustments, especially when multiple parameters were modified simultaneously (P3). 

To reduce this burden, participants advocated for several improvements:
\begin{itemize}
    \item \textbf{User profiles:} Nine participants suggested a user profiling system that saves and automatically applies preferred configurations across videos. P6 suggested a browser-integrated solution, such as Chrome extensions.
    \item \textbf{Preview and comparisons:} Seven participants wanted to preview how captions would change before committing to a new setting. Building on this, P3 proposed a comparison view to toggle between the current and previous configurations to better evaluate the difference.
\end{itemize}

\subsubsection{Balancing Information Richness and Caption Readability}
\label{sec:balance-richness}
Participants expressed a recurring tension between wanting rich, expressive captions and the cognitive effort required to process them. Eight participants noted that dense or fast-moving captions made it difficult to follow on-screen visuals. P6 described highly detailed captions in fast-paced scenes as \textit{``overwhelming''} as they often disappeared before she could finish reading them. P9 also highlighted this issue, stating that rapid caption changes posed particular difficulties for older viewers. To this end, P11 emphasized the importance of timing, noting that captions must remain on screen long enough to be fully read.

Five participants also noted that captions sometimes repeated information already apparent from the visuals. For example, P2 felt the caption \texttt{[Kristoff Panting with Short Breaths]} was redundant when the character was \textit{``visibly panting.''} P6 shared a similar view: \textit{``I really don't care for captions that describe actions I could already see happening.''}

\subsubsection{Context-Sensitive Preferences}
\label{sec:context-pref-findings}
Participants unanimously emphasized that caption preferences are highly context-dependent, shaped by content type, viewing intent, and the specific scene.
\paragraph{Content Types and Viewing Intentions.} All twelve participants expressed different preferences based on the type of video. P2 noted that source-focused captions \textit{``make sense for documentaries,''} while adjusting \textsc{Expressiveness} ``helps with a more cinematic experience.'' Six participants stated that their preferences would change depending on their mood or the purpose for watching. For example, P3 noted that after a long day, DHH children might experience listening fatigue and prefer simpler captions to avoid being overwhelmed.

\paragraph{Scene-Level Granularity.} Preferences also varied within a single video. P2 provided an example of a nature documentary, suggesting that slower scenes, such as \textit{``lions sleeping and grazing around,''} need minimal captions, while dramatic ones, such as \textit{``zebras being hunted by lions,''} call for more elaborate descriptions. He also stressed that off-screen sounds particularly benefit from richer descriptions, pointing to a sword sound in \textit{Frozen} that occurred off-camera: \textit{``This is where detailed captions become most valuable, as viewers could not see the source of the sound.''} This led several participants to desire more granular, context-aware controls instead of applying global settings. For example, P2 proposed segmenting the video based on scenes, with customization tailored to each section.

\subsubsection{Nuanced Perspectives on Sound Representation}
Participants offered nuanced perspectives on the three sound representation modes, emphasizing that each had advantages in specific contexts.
\paragraph{Onomatopoeia.} This method received mixed reactions. P2 appreciated that sound-mimicking words like ``swish'' for a sword helped in understanding the acoustic experience: \textit{``the word sounds like an action.''} P8 similarly associated onomatopoeia with comic books, where sounds were often represented visually through styled texts. However, five participants expressed reservations. P10 found expressions like ``swing'' and ``swoosh'' ambiguous, unsure whether they referred to a sword or another object. P11, who identifies as Deaf, noted a deeper challenge: \textit{``As a Deaf person, sometimes we might not understand the phonetic aspect of it,''} emphasizing that such representations can be inaccessible for those unfamiliar with \textit{``how sounds are supposed to sound.''}

\paragraph{Sensory Quality-Focused Descriptions.} This approach helped participants experience abstract or difficult-to-hear sounds more effectively. For example, we noted P2's experience with the ``dolphin whistle'' sound:
\begin{quote}
    \textit{Here it says ``Dolphin Whistle,'' but I don’t know what a whistle sounds like... [P2 selected ‘Sensory Quality-Focused’ in Sound Representation Options]... Ah, okay, here you go! The ``high-pitched squeak''! That’s what I am looking for.}
\end{quote} 
Similarly, P6 shared that sensory descriptors helped her comprehend high-pitch sounds that are typically missed, even with hearing aids.

\paragraph{Source-Focused Descriptions.} Source-focused descriptions, which provide more detail on sound sources, were seen as helpful for establishing clear context. P10 preferred this approach because it provided \textit{``enough critical details to see what’s going on from the sound and what’s happening in the scene.''}. However, some participants felt that the source alone was not always sufficient. P8 noted: ``I see `Sword Drawing,' but I don’t hear the sound associated with the sword drawing. So I’m thinking, what does it sound like?'' This sentiment highlights a desire for representations that combine both source and sensory cues.

Together, these perspectives suggest that no single representation method is universally preferred; rather, their effectiveness depends on the type of sound, the viewer’s hearing background, and the context in which the sound occurs.

\subsubsection{Concerns Around AI Interpretation and Consistency}
Many participants expressed concerns that AI-generated captions might be overly interpretive, potentially undermining their ability to form an independent understanding of the content. For example, seven noted discomforts with captions that imposed meanings rather than objectively describing sounds. As P8 explained: \textit{``The movie is thinking for me and I’m not having any thoughts or feelings about how this information is communicated.''} She further emphasized that \textit{``part of the movie experience is interpreting the content yourself.''}

P3 voiced a similar concern when encountering the caption \texttt{[Dolphins splash, full of vitality]} at an \textsc{Expressiveness} setting of 7. She questioned whether the caption was offering a factual description or an interpretation:
\begin{quote}
    \textit{They do seem to be happy right now, but maybe that’s just a matter of interpretation, right? I think high expressiveness may be bringing in way too many details that we may not need. We need to be free to make our own interpretation.}
\end{quote}
P8 echoed this with a more pointed critique of the caption \texttt{[Warm purr]} after selecting \textsc{(Level of Detail = 3, Expressiveness = 6)}, asking: \textit{``Here it says `the purr is warm,' but the question is, are you going to trust that?''}

Beyond interpretive overreach, participants also noted inconsistencies in how the system handled similar sounds across different contexts. P2 pointed out that the identical source captions like \texttt{[Dolphin whistles]} were sometimes transformed differently, depending on where they appeared in the video:
\begin{quote}
    \textit{I see the ``high-pitch'' description did not get picked up for the other `Dolphin Whistles.' If they are the same sound, you should ideally hope for it to be captioned the same way.}
\end{quote}
P11 expressed a broader concern about the reliability of AI-generated captions, stating: \textit{``You don’t want the AI to go off. I think we will need someone in the middle to help, tweak, and modify the words, making sure they match the intention.''}

These reflections underscore the importance of preserving DHH viewers' interpretive agency, ensuring semantic consistency, and providing mechanisms for human oversight in AI-mediated caption customization processes.

\subsubsection{Cultural and Linguistic Influences on Caption Needs}
Participants' linguistic and cultural backgrounds shaped their preferences. P7 noted that CapTune could support individuals with varying comprehension levels, particularly those who benefit from simplified language. However, P4 and P7 also cautioned that highly expressive captions could become inaccessible to some Deaf viewers, whose first language is not English: \textit{``Sometimes they don't have the vocabulary that includes these expressive words.''} (P4)

Diverse hearing histories also influenced needs. P6, who struggles with high-frequency sounds, appreciated how sensory-focused captions like \texttt{[High-pitched dolphin squeak]} filled perceptual gaps: \textit{``I would have never known what it sounded like if this description had not come up.''} Similarly, P4, who uses hearing aids, explained:
\begin{quote}
    \textit{Nuanced elements, especially high-pitched ones like wind sounds, are harder for me to catch. If the system can categorize and present this kind of auditory information effectively, I believe it could be really helpful.}
\end{quote}

\subsubsection{Envisioned Applications and Use Cases}
Most participants (N=11) expressed interest in using CapTune for everyday media experiences. They envisioned it as a valuable tool for enhancing comprehension, emotional connection, and accessibility---particularly when captions could be adjusted to suit different moods, genres, and energy levels. However, one participant (P5) expressed reservations: \textit{``I wouldn’t necessarily use it in the form that it is now,''} suggesting that further refinement would be needed for routine adoption.

Beyond personal use, participants identified several broader application areas where CapTune could have a meaningful impact. A prominent theme was educational use, especially for DHH children. For example, P7 described how the system could support language learning by gradually increasing caption complexity: 
\begin{quote}
    \textit{For Deaf people who have children learning English, I can imagine the captions being kept at a basic level for them to be able to learn. And then as they grow, they can adapt the captions to match their abilities.}
\end{quote}
P8 recognized CapTune's potential for music-focused content, suggesting that captions using musical terminology could convey musical elements more effectively than generic labels like ``dramatic music.''

These imagined use cases reflect a strong desire for captioning systems that go beyond baseline accessibility and serve as tools for learning, cultural participation, and artistic interpretation.

\section{Discussion}
CapTune reframes non-speech captions as dynamic, co-authored experiences---constructed not solely by creators or viewers, but collaboratively negotiated through constrained AI transformations. Our findings demonstrate the value of this approach for enhancing media accessibility among DHH audiences while preserving creator intent. In this section, we reflect on the broader implications of our study, connecting them to prior work and outlining opportunities for future systems.


\subsection{Beyond Visual Styling: Personalizing the Language of Non-Speech Captions}


Traditional captioning systems adopt a one-size-fits-all approach, offering static caption tracks that fail to reflect the diverse ways DHH viewers interpret and engage with sound. Prior work on enhancing non-speech information (NSI) in captions has focused primarily on visual augmentations---such as font styling \cite{de_lacerda_pataca_caption_2024, lee_emotive_2007}, background color \cite{may_towards_2024, vy_using_2010}, or animated overlays \cite{wang_visualizing_2017, alonzo_beyond_2022}---while leaving the textual content itself unchanged \cite{udo_rogue_2010}.

CapTune challenges this paradigm by enabling real-time, viewer-driven transformation of the caption language itself, allowing DHH viewers to adjust how non-speech sounds are described, how much detail is conveyed, and the tone of expression---from neutral to more creative or evocative phrasing. Using this tool, DHH viewers can adapt non-speech captions to better suit their preferences, needs, and viewing contexts. For example, participants shared that richly descriptive captions deepened their emotional connection to the content (Section~\ref{sec:viewer-ux}). This approach also resonates with the concept of \textit{access intimacy} \cite{noauthor_access_2011}, where accessibility feels intuitive and individually attuned. Rather than treating captions as purely functional, we position them as expressive cinematic elements, echoing the creative experimentation seen in stylized media captions (e.g., Stranger Things) \cite{salazar_wet_2022, alepa_stranger_2022, bitran_meet_2022}, while offering individual customization to support broader accessibility.


\subsection{Balancing Creator Intent and Viewer Agency}

Our findings underscore a longstanding tension in accessible media design: the need to balance creators’ narrative intentions with DHH viewers’ interpretive agency \cite{zdenek_reading_2015, bitran_meet_2022}. This is particularly relevant for non-speech captions, where interpretive language (e.g., “ominous hum,” “sarcastically”) can shape a viewer’s emotional or narrative perception. CapTune addresses this tension through an “anchored transformation” model, where creators define the upper and lower bounds of acceptable caption variation, thereby constraining the behavior of the generative model. This extends constrained text generation frameworks \cite{kumar2022gradient} into accessibility contexts. The approach strikes a middle ground between rigid standardization (e.g., captioning guidelines \cite{noauthor_captioning_2025, initiative_wai_captionssubtitles_2025}) and the risks of open-ended AI outputs, which may produce misleading or over-interpretive results \cite{ji_ai_2024, shen_towards_2024}.

In our study, creators appreciated retaining creative control through anchor setting, while DHH viewers valued the ability to personalize content. However, some viewers expressed discomfort with expressive captions that felt too interpretive, describing them as intrusive or restrictive to their own meaning-making. These concerns reflect broader critiques of AI systems that mediate user experience in ways that may obscure authorship or intent \cite{mankoff_disability_2010, alper_giving_2017, ouyang2022training, bennett_promise_2019}. Future work should support more granular viewer control over interpretive range, provide visible distinctions between objective and inferred content, and incorporate feedback mechanisms that help tune system behavior to user expectations.

\subsection{Context-Aware Caption Adaptation for Situated Preferences}

Caption preferences are not static. Our findings show that they shift depending on genre, scene content, and viewing goals. For example, DHH participants favored concise captions in documentaries but preferred richer and more expressive captions for ``cinematic experiences'' such as Disney movies (Section~\ref{sec:context-pref-findings}). This builds on prior research showing that DHH viewers prefer different caption styles for different genres \cite{alonzo_beyond_2022, may_enhancing_2023}. Even within a single video, preferences varied by scene: action-heavy or fast-paced moments required simpler captions to minimize reading effort, while off-screen or ambiguous sounds benefited from more descriptive phrasing to aid interpretation. This context-sensitivity aligns with prior work by Cambra et al. \cite{cambra_how_2010}, who found that deaf adolescents tend to shift their interpretations of scenes based on available cues---relying more on visual actions in fast-paced segments and on captions to understand character intentions.

CapTune addresses this partly through genre alignment and the use of localized scene context via audio-visual language models. Future work could expand on this by incorporating viewer context---such as social setting, attention level, or fatigue---and by developing AI agents that detect narrative shifts or scene transitions to dynamically adjust caption tone and density.




\subsection{Cognitive Load and Processing Constraints}

Several DHH participants described a core trade-off between information richness and cognitive load. Richer captions could enhance engagement and emotional clarity, but sometimes compete with visual processing, particularly in fast-paced or complex scenes (Section~\ref{sec:balance-richness}). This echoes previous findings on the split-attention effect in captioned content \cite{rashid_dancing_2008} and broader concerns around ``listening fatigue'' \cite{noauthor_listening_2025}.

Participants also noted that caption redundancy—e.g., describing visible actions like \texttt{[Kristoff Running]}—could distract or break immersion, while captions that covered off-screen or hard-to-visualize sounds (e.g., \texttt{[Dolphins' High-Pitched Whistle]}) were especially appreciated. Linguistic complexity was another barrier: ASL users or those with limited English proficiency flagged highly expressive or figurative language as inaccessible \cite{hoffmeister_acquiring_2014}. Viewers also expressed diverse preferences for sound representation such as onomatopoeia, sensory descriptors, or source-based cues—reflecting varied conceptualizations of sound within DHH culture \cite{rosen_representations_2007, holcomb_introduction_2013}.

These insights suggest the need for adaptable captioning strategies that allow users to toggle between modes of representation, simplify vocabulary, or preview caption styles to strike a balance between detail and readability.

\subsection{Creator–AI Collaboration and Personalization Workflows}

CapTune models a collaborative workflow where creators define boundaries and AI handles transformation. While creators valued this control and the efficiency of automation, they emphasized the need for ``debuggability''  \cite{he_err_2024}---the ability to trace and refine how AI-generated captions evolve (Section~\ref{sec:ct-utility}). This call for transparency reflects broader findings in HCI that emphasize the role of explainability in fostering user trust and control in creative systems \cite{ehsan_automated_2019}. DHH Viewers also expressed a desire for smoother, less labor-intensive personalization, as tuning parameters without clear feedback or persistent profiles felt burdensome. These challenges mirror longstanding accessibility concerns about personalization overhead and reinforce the need for adaptive systems that learn user preferences and reduce friction over time \cite{gajos_automatically_2010}.

Participants also highlighted practical needs for real-world adoption: creators wanted integration with professional workflows (e.g., DaVinci Resolve \cite{noauthor_davinci_2025}), while DHH viewers requested presets, previews, and context-aware caption switching. Together, these insights point to a future where personalization and professional workflows co-evolve to support inclusive captioning at scale. Collectively, these insights underscore the importance of AI systems that deliver high-quality outputs while accommodating the workflows and cognitive demands of diverse users.

\subsection{Toward Adaptive and Personalized Captioning Systems for Non-Speech Information}

Drawing from our findings, we identify five key design directions for future captioning systems:

\begin{enumerate}
    \item \textbf{Context-Aware, Granular Adaptations:} Enable captioning systems to adapt dynamically to genre, tone, pacing, and narrative structure, rather than applying uniform transformations across the entire video.
    \item \textbf{User Modeling with Preference Retention:} Reduce user burden through profiles and interaction history that help systems remember and apply user preferences.
    \item \textbf{Explainable Transformations:} Make caption changes transparent to both creators and viewers by exposing transformation logic and underlying decisions.
    \item \textbf{Cultural and Linguistic Adaptation:} Incorporate viewers’ linguistic proficiency and cultural context by enabling tailored vocabulary, idioms, and tone---especially for users who use ASL or approach English as a second language.
    
    \item \textbf{Semantically Aligned, User-Controlled Representations:} Recommend captioning styles based on sound characteristics and narrative context, while preserving user control to override or adjust these choices.
\end{enumerate}

\subsection{Limitations and Future Work}

While our results are promising, several limitations remain. First, the four customization parameters in CapTune—though grounded in our findings—represent a simplification of the rich, fluid space of DHH preferences. A linear interpolation model may not fully capture shifting needs across scenes, genres, or emotional contexts.

Second, our evaluation focused on short-form content, with clips ranging from 2 to 8 minutes in length. Longer content, such as full-length films or episodic television, introduces additional challenges related to narrative consistency, viewer fatigue, and cumulative comprehension—factors that our current study does not yet address.

Third, the system relies on the GPT-4o model, which, despite the guardrails we implemented, can still produce inconsistent or semantically inaccurate outputs. Future work should explore the use of alternative foundation models, such as Claude \cite{noauthor_meet_2025} or Gemini \cite{noauthor_gemini_2025}, and investigate fine-tuning or hybrid pipelines that combine LLMs with post-editing tools or human-in-the-loop validation for improved reliability.

Finally, our evaluation methodology focused on qualitative feedback from creators and DHH participants. While these insights offer depth and a user-centered perspective, they do not provide systematic metrics of caption transformation accuracy or consistency. Future work could incorporate large-scale quantitative evaluations using human raters to assess caption quality across diverse video datasets.

\section{Conclusion}
Current approaches to non-speech captions often adopt a one-size-fits-all model, overlooking the diverse preferences of DHH viewers. We introduced CapTune, a system that supports customizable non-speech captions through creator-defined transformation boundaries and viewer-facing controls. Our evaluations showed that the system enhanced narrative engagement for DHH viewers while preserving creators’ creative intent. By treating captions as adaptable and co-authored, CapTune points to a new direction for accessible media—one that embraces personalization, creative flexibility, and cultural nuance in support of more inclusive viewing experiences.

\bibliographystyle{ACM-Reference-Format}
\bibliography{references, references-2}

\clearpage

\appendix

\begin{table*}[t]
  \centering
  \caption{Secondary Concepts from Content Analysis}
  \label{tab:secondary_concepts}
  \renewcommand{\arraystretch}{1.3} 
  \begin{tabular}{@{}clp{9cm}@{}}
    \toprule
    \textbf{No.} & \textbf{Concept} & \textbf{Brief Description} \\
    \midrule
    1  & Sound Perception        & How DHH viewers conceptualize, interpret, and value sound information based on individual hearing histories \\
    2  & Brevity/Detail Balance  & Finding the optimal level of descriptive detail without overwhelming the viewer \\
    3  & Relevance Filtering     & Prioritizing sounds based on narrative importance and plot significance \\
    4  & Visual Attention        & Managing the competition between reading captions and viewing visual content \\
    5  & Reading Speed           & Considering the time needed to process caption text relative to visual scenes \\
    6  & Info Prioritization     & Determining which sounds deserve emphasis in limited caption space \\
    7  & Mood Conveyance         & Using captions to communicate emotional tone and atmosphere \\
    8  & Vocabulary Choices      & Selection of appropriate terminology that resonates with DHH viewers \\
    9  & Formatting Conventions  & Adherence to or deviation from standard captioning practices \\
    10 & Content/Genre Alignment & Matching caption style to content type (e.g., documentaries vs. movies) \\
    11 & Plot Significance       & Identifying sounds crucial to narrative understanding \\
    12 & Enhancing Storytelling  & Using captions to support narrative engagement and immersion \\
    13 & Depiction Method        & Approaches to representing sound (e.g., onomatopoeia vs. description) \\
    14 & Cultural Diversity      & Accounting for varied cultural interpretations of sound across DHH communities \\
    15 & Source Identification   & Clarifying the origin of sounds, especially for off-screen audio \\
    16 & Enjoyment/Immersion     & Supporting entertainment value and cinematic experience \\
    17 & Contextual Awareness    & Providing sound information that considers scene context \\
    18 & Personal Background     & Acknowledging how individual experiences shape caption preferences \\
    \bottomrule
  \end{tabular}
\end{table*}

\begin{table*}[t]
  \centering
  \caption{Self-reported demographics of study participants in the Creator Tool evaluation, including years of video creation experience and types of videos they have produced.}
  \label{tab:creator-demographics}
  \renewcommand{\arraystretch}{1.3} 
  \begin{tabular}{@{}ccccp{7cm}@{}}
    \toprule
    \textbf{ID} & \textbf{Age} & \textbf{Gender} & \textbf{Years of Experience} & \textbf{Types of Videos Created} \\
    \midrule
    C1  & 24 & Male & 4 & Tech demos  \\
    C2  & 29 & Male & 2 & Short movies  \\
    C3  & 25 & Male & 7 & Education, music, tech demos, gaming  \\
    C4  & 26 & Female & 8 & Education, tech demos, lifestyle  \\
    C5  & 22 & Male & 8 & Technology reviews  \\
    C6  & 38 & Female & 4 & Professioal ads, independent films  \\
    C7  & 37 & Female & 12 & Travel blogs, lifestyle  \\
    \bottomrule
  \end{tabular}
\end{table*}

\begin{table*}[t]
  \centering
  \caption{Self-reported demographics of study participants in the Viewer Client evaluation, including their current sentiment toward non-speech captions.}
  \label{tab:viewer-demographics}
  \renewcommand{\arraystretch}{1.3} 
  \begin{tabular}{llllllll}
    \toprule
    \textbf{ID} & \textbf{Gender} & \textbf{Age} & \textbf{Identity} & \textbf{Hearing loss} & \textbf{Primary Language} & \textbf{Current Sentiment} \\
    \midrule
    P1 & Female & 49 & Deaf  & Profound & ASL, English  & Moderately satisfied \\
    P2 & Male & 25 & Hard of Hearing  & Severe (L), No (R) & English & Moderately satisfied \\
    P3 & Female & 26 & Hard of Hearing  & Severe & English  & Slightly dissatisfied \\
    P4 & Female & 26 & Hard of Hearing  & Profound & English & Slightly dissatisfied \\
    P5 & Male & 23 & Deaf  & Profound & English  & Slightly dissatisfied \\
    P6 & Female & 37 & Hard of Hearing  & Profound & English  & Slightly dissatisfied \\
    P7 & Female & 29 & Deaf  & Profound & ASL  & Not at all satisfied \\
    P8 & Female & 31 & deaf  & Not disclosed & English & Slightly dissatisfied \\
    P9 & Male & 67 & Hard of Hearing  & Severe & English  & Not at all satisfied \\
    P10 & Male & 24 & deaf  & Profound & English  & Not at all satisfied \\
    P11 & Female & 61 & Deaf & Profound & ASL, English  & Very satisfied \\
    P12 & Male & 62 & Deaf  & Profound & English & Slightly dissatisfied \\
    \bottomrule
  \end{tabular}
\end{table*}

        

\end{document}